\documentclass[sigconf]{acmart}

\AtBeginDocument{%
  }

\copyrightyear{2026}
\acmYear{2026}
\setcopyright{cc}
\setcctype{by}
\acmConference[GLSVLSI '26]{Great Lakes Symposium on VLSI 2026}{June 22--24, 2026}{Canandaigua, NY, USA}
\acmBooktitle{Great Lakes Symposium on VLSI 2026 (GLSVLSI '26), June 22--24, 2026, Canandaigua, NY, USA}
\acmDOI{10.1145/3787109.3816396}
\acmISBN{979-8-4007-2431-2/2026/06}

\setcopyright{none}   

\begin{document}


\title{Lightweight Cross-Device Sleep Tracking on the WeBe Wearable Platform}



\author{Wei Shao}
\affiliation{%
  \institution{University of California, Davis}
  \city{Davis}
  \state{California}
  \country{USA}
}
\email{wayshao@ucdavis.edu}

\author{Ehsan Kourkchi}
\affiliation{%
  \institution{University of California, Davis}
  \city{Davis}
  \state{California}
  \country{USA}
}
\email{ekay@ucdavis.edu}

\author{Krishi Prashant Shah}
\affiliation{%
  \institution{University of California, Davis}
  \city{Davis}
  \state{California}
  \country{USA}
}
\email{krishah@ucdavis.edu}

\author{Zequan Liang}
\affiliation{%
  \institution{University of California, Davis}
  \city{Davis}
  \state{California}
  \country{USA}
}
\email{zqliang@ucdavis.edu}

\author{Setareh Rafatirad}
\affiliation{%
  \institution{University of California, Davis}
  \city{Davis}
  \state{California}
  \country{USA}
}
\email{srafatirad@ucdavis.edu}

\author{Houman Homayoun}
\affiliation{%
  \institution{University of California, Davis}
  \city{Davis}
  \state{California}
  \country{USA}
}
\email{hhomayoun@ucdavis.edu}

\renewcommand{\shortauthors}{Shao et al.}

\begin{abstract}
Wearable devices are widely used for continuous health monitoring, yet reliable sleep tracking on emerging platforms remains underexplored due to reliance on proprietary algorithms and device-specific activity representations. We present a lightweight and reproducible sleep tracking pipeline that operates directly on raw accelerometer signals. The method converts data into epoch-level activity features, applies temporal smoothing and normalized scoring, and performs sleep/wake classification using a globally calibrated threshold. We calibrate the model on the Multilevel Monitoring of Activity and Sleep in Healthy People (MMASH) dataset and evaluate it in a cross-device study using the WeBe wearable platform and a commercial ActiGraph device. On MMASH, the method achieves a mean absolute error of 41.6 minutes in Total Sleep Time (TST), with onset and offset errors of 6.3 and 7.4 minutes. On real-world WeBe data from three participants across five sessions, it achieves a mean TST error of 27.4 minutes and onset and offset errors of 13.9 and 8.0 minutes. In contrast, a commercial ActiGraph pipeline shows larger discrepancies relative to ground truth. These results demonstrate accurate and generalizable sleep tracking using a simple and reproducible pipeline.
\end{abstract}


\begin{teaserfigure}
\centering
  \includegraphics[width=0.8\textwidth]{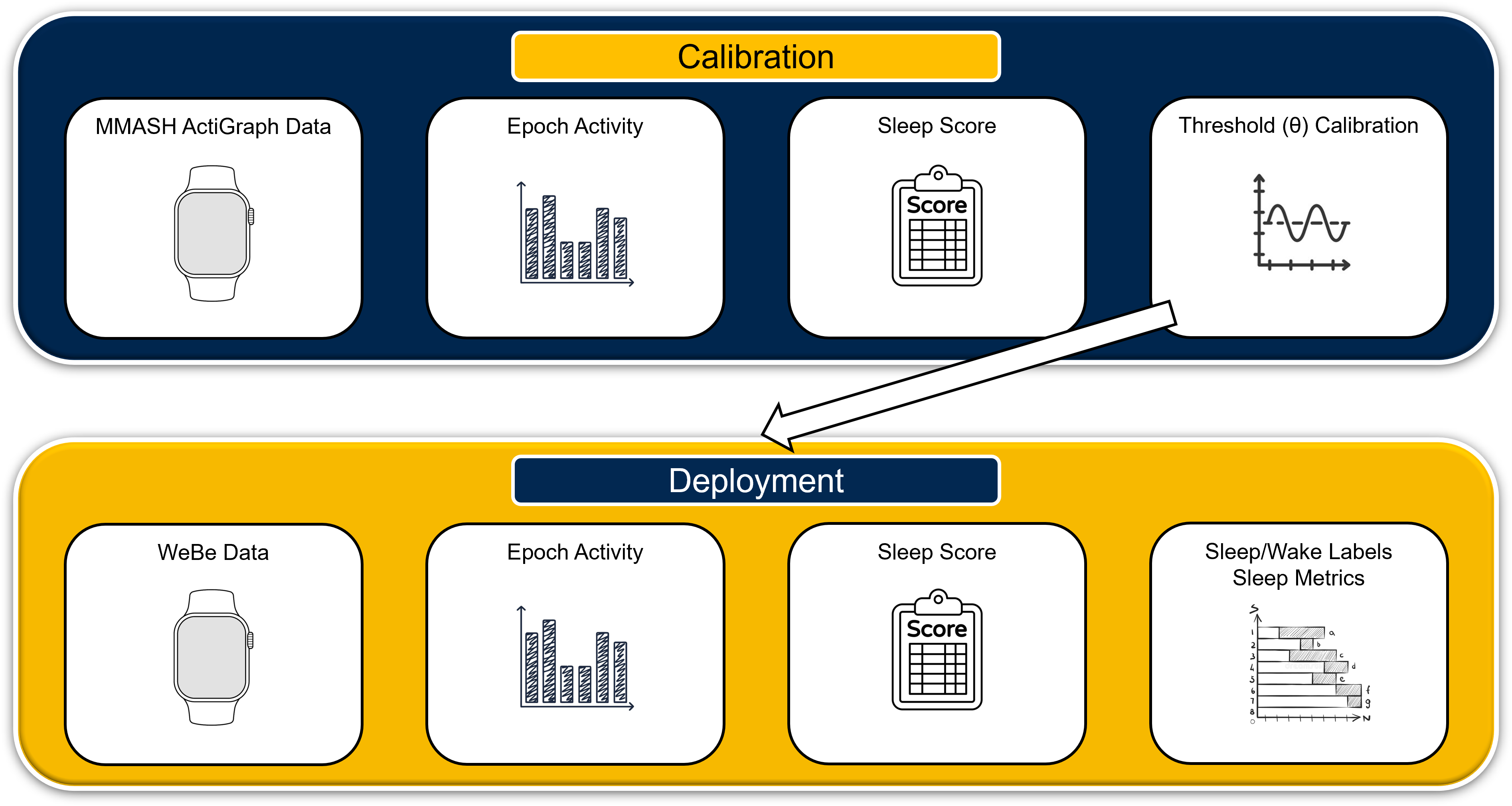}
  \caption{Overview of the proposed sleep tracking pipeline.}
  \Description{Overview of the proposed sleep tracking pipeline.}
  \label{fig:teaser}
\end{teaserfigure}


\maketitle

\section{Introduction}

Sleep plays a critical role in human health, affecting cognitive performance, cardiovascular function, and overall well-being~\cite{atkinson2007relationships}. As a result, continuous sleep monitoring has become an essential component of modern wearable systems~\cite{de2019wearable}. Commercial devices based on ActiGraph estimate sleep by analyzing wrist-worn accelerometer signals and have been widely adopted due to their non-intrusive and low-power operation~\cite{acebo2006actigraphy}. However, existing solutions often rely on proprietary algorithms and device-specific activity representations, limiting transparency, reproducibility, and adaptability to emerging wearable platforms.


Recent research has explored custom wearable platforms that provide richer sensing capabilities and greater flexibility for algorithm development. In particular, the WeBe Band has been introduced as a research-oriented wearable device supporting multimodal physiological sensing, including accelerometry, photoplethysmography (PPG), electrodermal activity (EDA), and skin temperature, within a unified and programmable framework~\cite{zhang2024introducing,fang2024validation}. It has been used in a range of studies, including physiological monitoring and authentication, demonstrating its potential as a versatile platform for wearable systems research~\cite{shao2025know,liang2025rapid,liang2025generalizable,shao2025self}. The platform enables transparent data processing pipelines that operate directly on raw signals, supporting flexible experimentation across sensing modalities and application domains. In addition, WeBe supports the development and execution of lightweight models on embedded hardware, enabling efficient on-device processing and bridging the gap between algorithm design and practical deployment. Despite these capabilities, sleep tracking functionality on such platforms remains underexplored, particularly in a manner that is both reproducible and comparable to established ActiGraph-based methods.

In this work, we take a step toward practical and transparent sleep tracking on a custom wearable platform. We develop a lightweight, data-driven sleep estimation pipeline that operates directly on raw accelerometer signals without relying on proprietary activity counts~\cite{freedson2005calibration,matthew2005calibration}. Our approach uses epoch-based activity aggregation and a normalized activity scoring mechanism to classify sleep and wake states. We further adopt a threshold-based strategy calibrated on a public dataset, enabling the method to generalize across users and devices without per-user tuning.

To ensure robustness and generalizability, we calibrate our model using the publicly available Multilevel Monitoring of Activity and Sleep in Healthy People (MMASH) dataset, which provides ActiGraph data along with annotated sleep metrics~\cite{PhysioNet-mmash-1.0.0}. The calibrated model is then evaluated across multiple users without further parameter adjustment. Our evaluation shows that the proposed method achieves accurate estimation of sleep timing, with mean sleep onset and offset errors of 6.3 and 7.4 minutes, respectively. Total sleep time (TST) is estimated with a mean absolute error of 41.6 minutes, comparable to typical ActiGraph-based approaches. While Wake After Sleep Onset (WASO) remains challenging due to the difficulty of detecting brief awakenings from motion signals alone, the method provides reliable identification of primary sleep periods.

Building on this foundation, we apply the calibrated model to data collected from the WeBe Band in a co-wear setting with a commercial ActiGraph device. This enables cross-device validation and provides insight into the feasibility of deploying sleep tracking functionality on emerging wearable platforms.

It is worth noting that the current study explores simplicity, interpretability, and device-agnostic performance as primary design objectives, rather than optimizing for maximal accuracy.

In summary, this work makes the following contributions:
\begin{itemize}
    \item We develop a lightweight and reproducible sleep tracking pipeline operating directly on raw accelerometer data without proprietary activity counts, and show that its simple computational structure makes it well-suited for on-device deployment on a wearable platform.
    \item We introduce a normalized activity scoring and threshold calibration approach using a public dataset, enabling cross-user and cross-device generalization.
    \item We provide a comprehensive evaluation on the MMASH dataset, demonstrating strong performance in sleep timing and competitive accuracy in sleep duration estimation.
    \item We demonstrate the applicability of the approach to the WeBe Band through cross-device validation with a commercial ActiGraph device.
\end{itemize}

\section{Methodology}
\label{sec:methodology}
\begin{figure}[t]
    \centering
    \includegraphics[width=0.35\columnwidth]{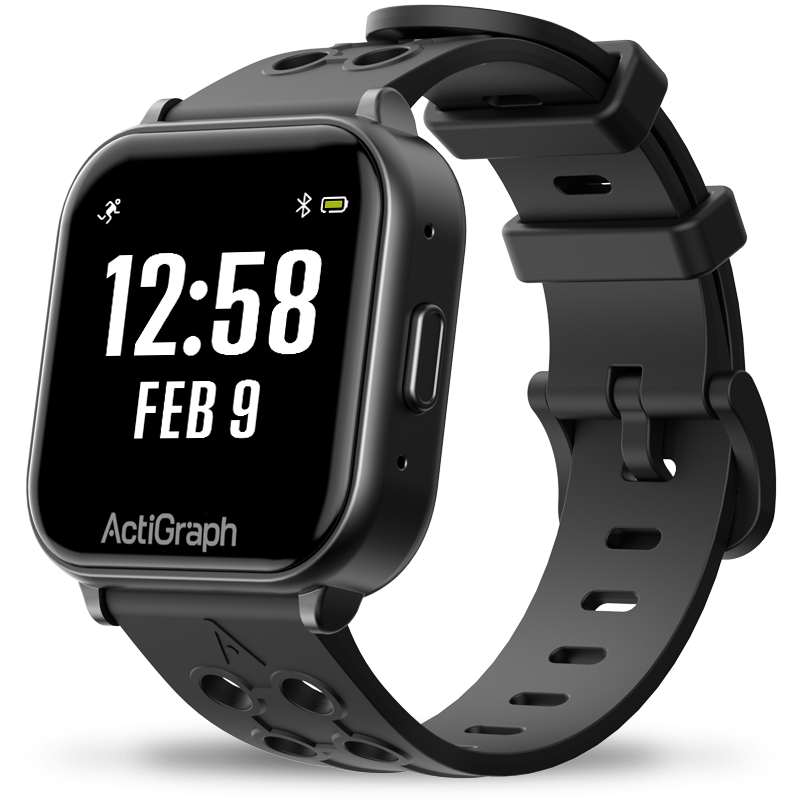}
    \hfill
    \includegraphics[width=0.55\columnwidth]{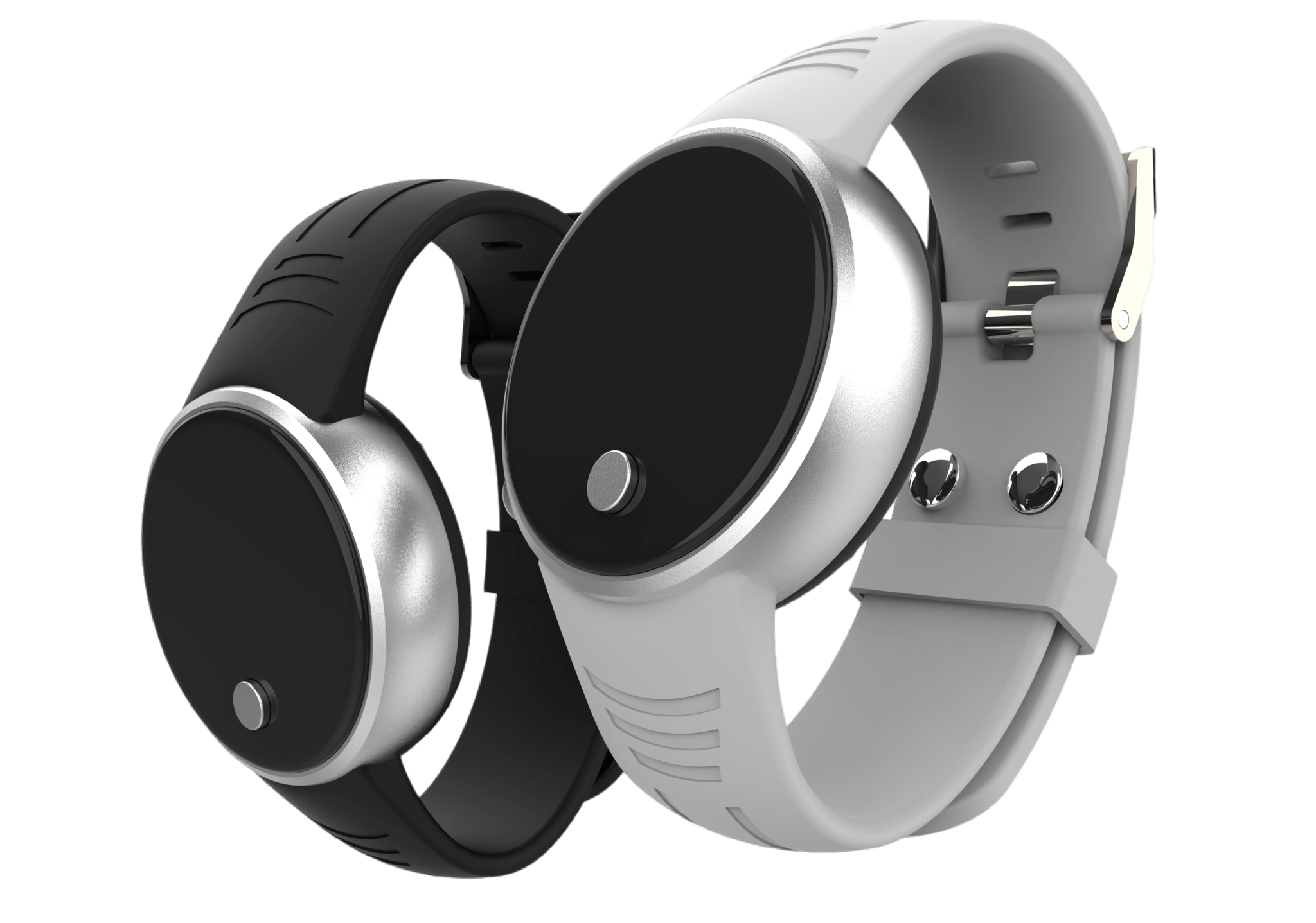}
    \caption{Wearable devices used in this study: (a) commercial ActiGraph LEAP~\cite{actigraph_web}, and (b) WeBe Band~\cite{webe_web}.}
    \Description{Wearable devices used in this study}
    \label{fig:devices}
\end{figure}

In this section, we describe our sleep tracking pipeline, which operates directly on raw accelerometer data and produces sleep metrics comparable to ActiGraph-based systems. Our design emphasizes on simplicity, adaptability, and robustness to device variation.

\subsection{Overview}

Our approach follows a three-stage pipeline: (1) epoch-based activity extraction from raw accelerometer signals, (2) sleep/wake classification using a normalized activity score, and (3) sleep metric computation within detected sleep periods. The overall workflow is illustrated as:

\begin{center}
\textit{Raw Accelerometer $\rightarrow$ Epoch Activity $\rightarrow$ Sleep Score $\rightarrow$ Sleep/Wake Labels $\rightarrow$ Sleep Metrics}
\end{center}

Our proposed pipeline relies only on fixed-size temporal filtering and simple arithmetic operations, with $O(N)$ complexity per epoch and small fixed memory buffers. In practice, this translates to a few hundred operations per epoch and negligible memory overhead, making the approach fit for real-time implementation on resource-scarce embedded devices.

\subsection{Epoch-Based Activity Extraction}

For each accelerometer sample, we compute the vector magnitude:
\begin{equation}
VM_i = \sqrt{x_i^2 + y_i^2 + z_i^2}.
\end{equation}

The continuous signal is segmented into fixed-length epochs of duration $T_e$ seconds. For each epoch $t$, we compute a count-like activity proxy by summing vector magnitudes:
\begin{equation}
A_t = \sum_{i=1}^{N_t} VM_i,
\end{equation}
where $N_t$ is the number of samples in epoch $t$.

\subsection{Temporal Smoothing and Sleep Score}

We smooth the epoch activity using a centered five-epoch moving average:
\begin{equation}
\tilde{A}_t = \frac{1}{5}\sum_{k=-2}^{2} A_{t+k}.
\end{equation}

We then compute a weighted contextual activity score:
\begin{equation}
S_t = \sum_{k=-3}^{3} w_k \tilde{A}_{t+k},
\end{equation}
where
\begin{equation}
\mathbf{w} = [0.04, 0.12, 0.20, 0.28, 0.20, 0.12, 0.04].
\end{equation}

The score is normalized using robust statistics:
\begin{equation}
\hat{S}_t = \frac{S_t - Q_{50}}{Q_{90} - Q_{50}},
\end{equation}
where $Q_{50}$ and $Q_{90}$ are the median and 90th percentile of $S_t$ over the recording.

\subsection{Sleep/Wake Classification}

Each epoch is classified using a calibrated threshold $\theta$:
\begin{equation}
y_t =
\begin{cases}
1, & \hat{S}_t < \theta,\\
0, & \hat{S}_t \geq \theta,
\end{cases}
\end{equation}
where $y_t=1$ denotes sleep and $y_t=0$ denotes wake. In our experiments, $\theta$ is calibrated on the MMASH dataset via grid search and fixed at $-0.05$ for all subsequent experiments.

\subsection{Sleep Period Detection}

To identify the primary sleep interval, we adopt a rule-based approach inspired by classical ActiGraph sleep–wake detection algorithms
such as Cole–Kripke and Sadeh~\cite{cole1992automatic,sadeh1994activity,tudor2014fully}. A sleep period begins when a sequence of consecutive epochs classified as sleep exceeds a predefined duration (bedtime definition), and ends when a sequence of consecutive wake epochs exceeds another duration (wake-time definition). In our implementation, we define sleep onset as at least 5 minutes of consecutive sleep epochs and sleep offset as at least 10 minutes of consecutive wake epochs. This approach ensures robust detection of sustained sleep periods while filtering out transient fluctuations.

\subsection{Sleep Metric Computation}

Within each detected sleep period, we compute standard sleep metrics:

\begin{itemize}
    \item \textbf{Sleep Onset:} the timestamp of the first epoch classified as sleep within the detected sleep period.
    \item \textbf{Sleep Offset:} the timestamp of the last epoch classified as sleep within the detected sleep period.
    \item \textbf{TST:} total duration of epochs classified as sleep.
    \item \textbf{WASO:} total duration of wake epochs occurring after sleep onset.
    \item \textbf{Sleep Efficiency:} ratio of TST to total time in bed, where time in bed is defined as the duration between the detected sleep onset and sleep offset.
\end{itemize}

To evaluate temporal accuracy, we compare predicted sleep onset and offset times with ground-truth annotations. The error is computed as the minimum absolute difference within a 24-hour cycle to account for overnight transitions.

\subsection{Cross-Dataset Calibration}

The classification threshold $\theta$ is calibrated using the MMASH dataset via a grid search over candidate values. Specifically, we evaluate thresholds in the range $[-0.4, 1.0]$ with uniform spacing, and for each candidate threshold, we compute prediction error across users.

The selection criterion is based on minimizing the mean absolute error of TST across all users and nights. Additional metrics, including WASO and sleep efficiency, are also monitored during calibration but are not used as the primary optimization objective.

In our experiments, the optimal threshold is found to be $\theta = -0.05$, which yields the lowest TST error on the MMASH dataset. This threshold is then fixed and applied unchanged to all subsequent datasets, including data collected from the WeBe Band.

\subsection{Edge Deployment and System-Level Considerations}

The proposed sleep tracking pipeline is designed with edge deployment in mind, targeting resource-constrained wearable devices such as the WeBe Band. The device is built on an nRF52840 system-on-chip integrating an Arm Cortex-M4 processor with Bluetooth Low Energy (BLE) connectivity, enabling both local computation and wireless communication~\cite{zhang2024introducing}. The computational requirements of the proposed pipeline are minimal, consisting primarily of vector magnitude computation, fixed-size temporal filtering, and threshold-based classification. These operations can be executed efficiently on embedded microcontrollers.

In the current prototype, data collected from the WeBe Band is transmitted to an external device for processing. However, the simplicity of the proposed pipeline makes it well-suited for future on-device deployment, enabling real-time sleep analysis without reliance on external computation resources. This design aligns with emerging trends in edge AI, where lightweight inference is performed directly on wearable devices.

In contrast to traditional ActiGraph workflows, where raw data is collected on-device and later transferred to a host system for post-hoc analysis using proprietary software, the proposed approach eliminates the need for offline processing pipelines when deployed on-device. This reduces system complexity and removes dependency on licensed software environments.

Given a sampling rate of 25 Hz and tri-axial accelerometer data (3 x 16-bit values), raw data collected during an 8-hour sleep period amounts to approximately 4 MB. While this volume is modest in terms of storage, traditional pipelines require exporting this data for external processing, introducing additional steps, latency, and manual synchronization. In contrast, an on-device implementation would require transmitting only compact summaries or sleep metrics, significantly reducing data movement and simplifying system operation.

From a system perspective, minimizing data transfer also improves privacy and security by keeping raw physiological signals on-device rather than transmitting them to external systems. Furthermore, reducing reliance on external processing enables continuous monitoring with immediate feedback, without requiring user intervention or periodic data synchronization.

The WeBe platform supports flexible deployment of data processing pipelines across edge and cloud environments~\cite{zhang2024introducing}. Our approach leverages this capability by enabling a lightweight implementation today, while providing a clear path toward future on-device deployment. These characteristics make the proposed method well-suited for edge AI-enabled wearable systems, where energy efficiency, latency, privacy, and usability are critical considerations.

The proposed pipeline maintains only small fixed-size temporal buffers for filtering and contextual scoring, requiring fewer than 20 epochs of temporary storage. At a 30-second epoch size, sleep inference is performed only once every epoch and consists primarily of vector magnitude computation, moving-average filtering, weighted summation, and threshold comparison. As a result, the computational cost is minimal and well within the capability of low-power embedded processors such as the Cortex-M4 used in the WeBe platform. These characteristics support the lightweight design objective of the proposed method and provide a practical path toward future real-time on-device deployment.

\section{Evaluation}

In this section, we evaluate the performance of the proposed sleep tracking pipeline using a public dataset and present cross-device validation results.

\subsection{Dataset and Experimental Setup}

We evaluate our method using the publicly available MMASH dataset. MMASH contains multi-day physiological and behavioral data collected from 22 healthy participants in free-living conditions. The dataset includes wrist-worn ActiGraph signals, heart rate measurements, and self-reported sleep annotations.

In this work, we use the ActiGraph component, which provides tri-axial accelerometer data sampled at 1 Hz. Each participant record includes day and time information, along with sleep annotations such as in-bed time, sleep onset, TST, WASO, and sleep efficiency.

We preprocess the raw accelerometer data by constructing timestamps from the provided day and time fields, followed by segmentation into fixed-length epochs of 30 seconds. For each epoch, we compute activity features as described in Section~\ref{sec:methodology}. Sleep classification is then performed using the normalized activity score with a fixed threshold $\theta = -0.05$, calibrated on the MMASH dataset.

Sleep annotations in MMASH may span across midnight; therefore, we align predicted and ground-truth sleep windows by explicitly handling overnight transitions. After preprocessing and alignment, we obtain valid sleep windows for 20 nights across 19 users, which are used for evaluation.

All parameters are fixed globally, and no per-user or per-night tuning is performed.

\subsection{Evaluation Metrics}

We evaluate the performance of our method using standard sleep metrics:

\begin{itemize}
    \item \textbf{TST:} total duration of epochs classified as sleep.
    \item \textbf{WASO:} total duration of wake epochs after sleep onset.
    \item \textbf{Sleep Efficiency:} ratio of TST to time in bed.
\end{itemize}

In addition, we evaluate temporal accuracy:

\begin{itemize}
    \item \textbf{Sleep Onset Error:} absolute difference between predicted and ground-truth sleep onset times.
    \item \textbf{Sleep Offset Error:} absolute difference between predicted and ground-truth sleep offset times.
\end{itemize}

Timing errors are computed as the minimum absolute difference within a 24-hour cycle to account for overnight transitions.

\subsection{Results on MMASH}

Table~\ref{tab:mmash_results} summarizes the performance of the proposed method across all evaluated nights. Our method achieves accurate estimation of sleep timing, with mean onset and offset errors of 6.3 and 7.4 minutes, respectively. The median errors are below one minute, indicating precise alignment for a large fraction of nights.

\begin{table}[t]
\centering
\caption{Sleep tracking performance on the MMASH dataset.}
\label{tab:mmash_results}
\begin{tabular}{lccc}
\toprule
\textbf{Metric} & \textbf{Mean} & \textbf{Median} & \textbf{Std} \\
\midrule
TST MAE (min) & 41.6 & 32.5 & 33.3 \\
WASO MAE (min) & 53.5 & 51.0 & 28.1 \\
Sleep Efficiency MAE (\%) & 12.1 & 9.8 & 5.9 \\
Sleep Onset Error (min) & 6.3 & 0.5 & 12.9 \\
Sleep Offset Error (min) & 7.4 & 0.0 & 17.1 \\
\bottomrule
\end{tabular}
\end{table}

The measured TST inaccuracy is in line with ranges published for accelerometer-based sleep estimation in free-living settings, where significant uncertainty can be introduced by user behavior variability and annotation quality. Crucially, the technique shows modest timing mistakes for sleep onset and offset, which in turn suggests that primary sleep intervals can be reliably identified even in cases when aggregate duration errors are larger.

\begin{figure}[t]
    \centering
    \includegraphics[width=\columnwidth]{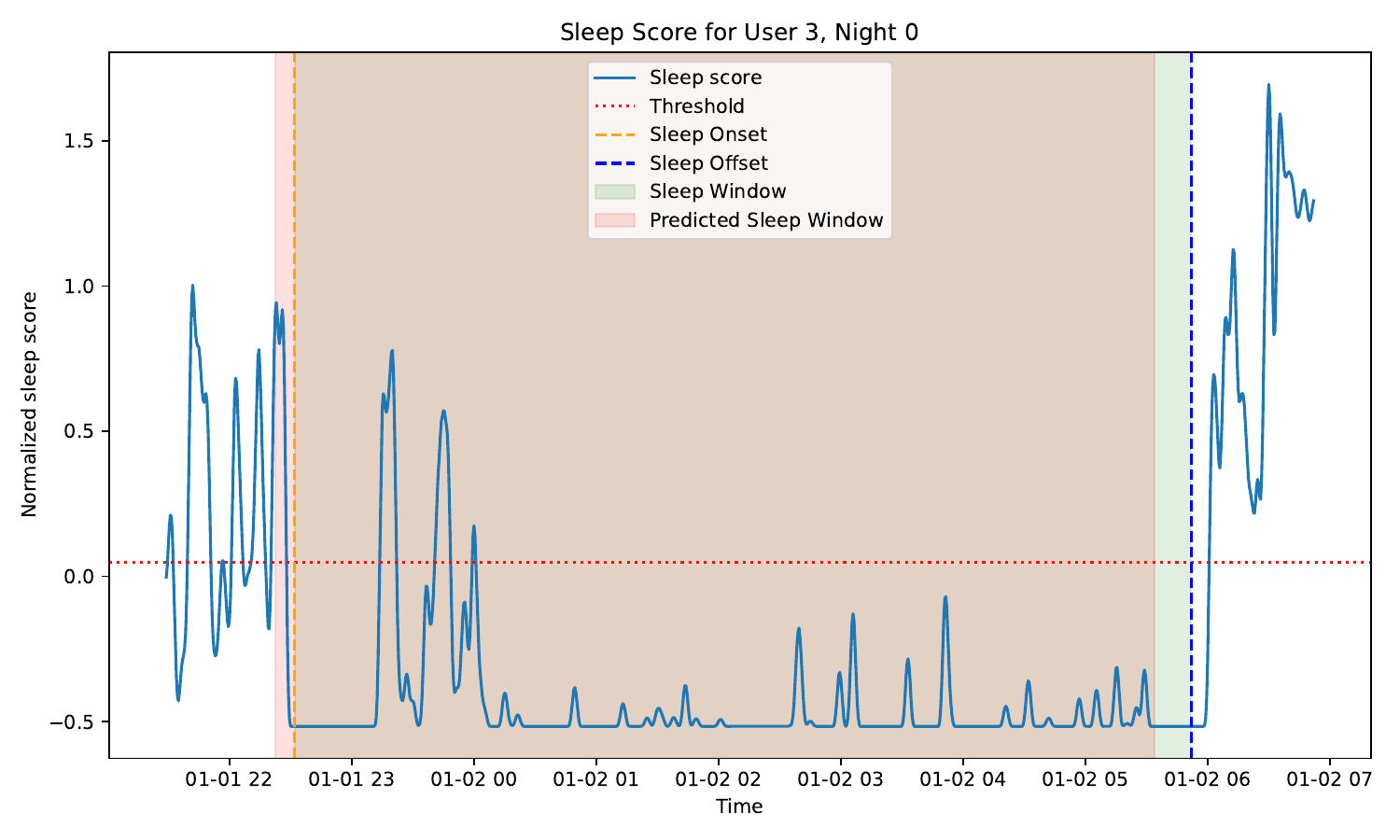}
    \caption{Example sleep score trajectory for a MMASH user.}
    \Description{Example sleep score}
    \label{fig:sleep_example}
\end{figure}

For sleep quantity, the method achieves a mean absolute error of 41.6 minutes for TST, which is within the range reported for ActiGraph-based approaches. However, WASO estimation remains challenging, with a mean error of 53.5 minutes. This is primarily due to the difficulty of detecting brief awakenings from accelerometer signals~\cite{marino2013measuring}.

Failure cases are primarily observed during low-motion wake periods, such as quiet rest or sedentary behavior, where accelerometer signals resemble sleep-like inactivity. In such scenarios, the method may overestimate sleep duration or underestimate WASO, reflecting a known limitation of motion-based sleep tracking approaches.

As illustrated in Fig.~\ref{fig:sleep_example}, the proposed method produces a normalized activity score that remains consistently below the threshold during sleep periods and aligns closely with the ground-truth sleep onset and offset times.

\subsection{Threshold Sensitivity}

\begin{figure}[t]
    \centering
    \includegraphics[width=\columnwidth]{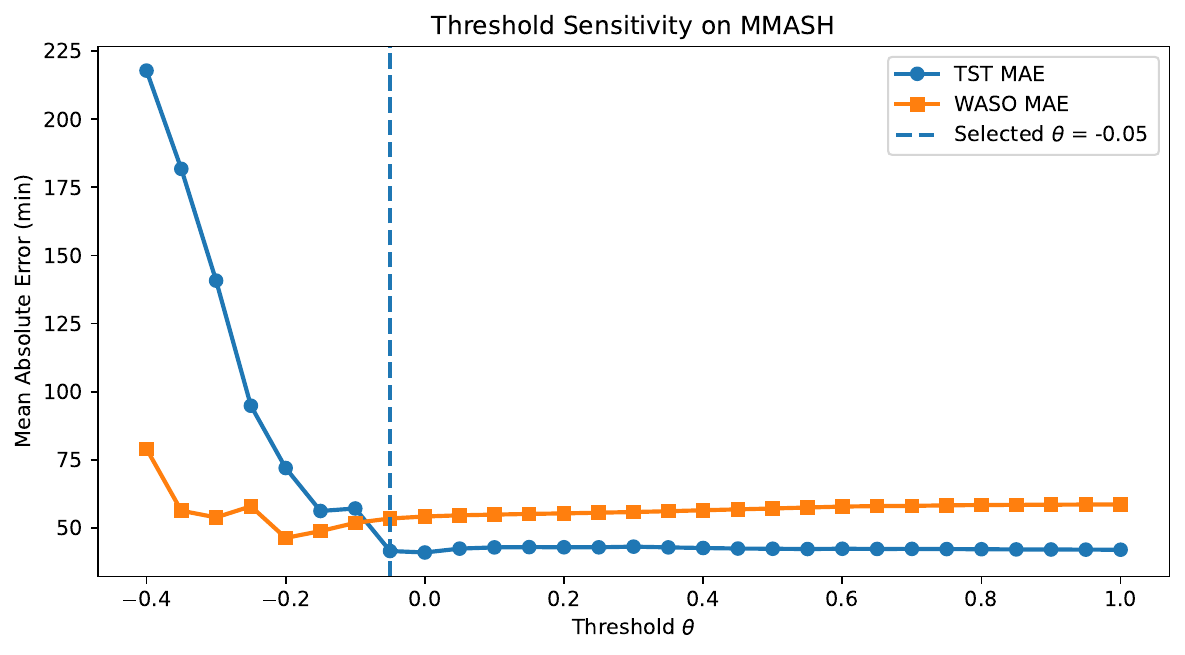}
    \caption{Threshold sensitivity on the MMASH dataset.}
    \Description{Threshold sensitivity}
    \label{fig:threshold_sensitivity}
\end{figure}

To evaluate the robustness of the calibrated threshold, we perform a grid search over candidate values in the range $[-0.4, 1.0]$ on the MMASH dataset. For each threshold, we compute the mean absolute error of TST and WASO across all valid user-night samples.

Figure~\ref{fig:threshold_sensitivity} shows the sensitivity of sleep tracking performance to the threshold value. The selected threshold $\theta=-0.05$ minimizes TST error, achieving a mean TST MAE of 41.6 minutes. Although WASO remains more challenging, the selected threshold provides a reasonable balance between sleep duration estimation and wake sensitivity. 

The performance curve shows a relatively stable region around the optimal threshold, indicating that small deviations from $\theta=-0.05$ do not significantly degrade performance. This suggests that the proposed method is robust to threshold selection and does not require precise fine-tuning. However, larger threshold values lead to rapid degradation in TST accuracy, highlighting the importance of proper calibration.

We therefore fix this threshold for all subsequent experiments, including the WeBe Band cross-device validation. This analysis confirms that the threshold is selected systematically rather than manually tuned.

\subsection{Cross-Device Validation on WeBe Band}

To evaluate the applicability of the proposed method in real-world settings, we conduct a cross-device validation study using data collected from the WeBe Band in a co-wear configuration with a commercial ActiGraph device.

In this study, three participants wore both devices simultaneously on the same wrist across five recording sessions. Data were collected in uncontrolled, free-living environments without any imposed routines, allowing participants to follow their natural daily and sleep behaviors. This setup reflects realistic usage conditions for wearable sleep monitoring.

We apply the proposed pipeline, calibrated on the MMASH dataset with a fixed threshold $\theta = -0.05$, directly to the WeBe Band accelerometer data without any additional tuning. Sleep metrics derived from the WeBe Band are compared against ground-truth sleep intervals obtained from participant annotations, while ActiLife outputs from the ActiGraph device are included as a reference for comparison.

Ground-truth sleep onset and offset times are recorded at minute-level resolution. Since the proposed method operates on fixed-length epochs, predicted timestamps are inherently quantized to epoch boundaries. Timing comparisons are therefore evaluated with respect to this resolution.


For all studies sleep sessions, our method produces sleep duration estimates that closely align with ground-truth observations. The method achieves a mean absolute error of 27.4 minutes in TST (median: 23.0 minutes). Sleep timing is estimated with mean onset and offset errors of 13.9 and 8.0 minutes, respectively, with standard deviations of 16.1 and 17.4 minutes.

\begin{table}[t]
\centering
\caption{Sleep tracking performance of the proposed method on WeBe Band (co-wear study).}
\label{tab:webe_results}
\begin{tabular}{lccc}
\toprule
\textbf{Metric} & \textbf{Mean} & \textbf{Median} & \textbf{Std} \\
\midrule
TST MAE (min) & 27.4 & 23.0 & 25.6 \\
Onset Error (min) & 13.9 & 11.8 & 16.1 \\
Offset Error (min) & 8.0 & 8.5 & 17.4 \\
\bottomrule
\end{tabular}
\end{table}


In contrast, ActiLife outputs exhibit inconsistencies in sleep duration and timing relative to ground truth under default processing settings. On average, ActiLife differs from ground-truth TST by approximately 50 minutes across sessions. These outputs are generated using the proprietary ActiGraph software without access to internal algorithm parameters, reflecting typical real-world usage. This is notably higher than the $\sim$28-minute error achieved by our proposed method. More pronounced deviations are observed in sleep timing; for example, in one representative session, ActiLife detects the sleep onset and offset approximately three hours earlier than the ground-truth interval, indicating a substantial temporal misalignment. These observations are qualitative and should not be interpreted as a controlled comparative evaluation.


The cross-device results highlight two key observations. First, the proposed method maintains consistent performance across both public (MMASH) and in-the-wild WeBe Band data using a single globally calibrated threshold. This suggests that the normalized activity representation effectively mitigates device- and user-specific variability, enabling practical deployment without per-user or per-device re-calibration.

Second, while ActiLife outputs exhibit moderate discrepancies in TST (on average approximately 50 minutes), more pronounced differences are observed in sleep timing. In particular, we observe a case where the detected sleep period is shifted by approximately three hours relative to the ground-truth interval. Such temporal misalignment can significantly affect derived sleep metrics and user interpretation, even when aggregate duration differences appear moderate.

These observations suggest that proprietary ActiGraph pipelines may be sensitive to variations in real-world activity patterns, especially in unconstrained environments where user behavior deviates from typical assumptions. In contrast, the proposed approach relies on a transparent and data-driven formulation based on normalized activity signals, which appears to generalize more reliably across datasets and devices.

Moreover, the limited number of participants and recording sessions restricts the statistical generalizability of these findings. Hence, any conclusions made here should be interpreted as preliminary evidence of feasibility rather than definitive performance, motivating larger-scale studies for more thorough validation.


\section{Conclusion}

In this work, we present a lightweight sleep tracking solution that operates directly on raw accelerometer signals without relying on proprietary activity counts. The proposed method combines epoch-based activity aggregation, normalized activity scoring, and a globally calibrated threshold to enable consistent sleep/wake classification across users and devices.

We evaluate the approach on the MMASH dataset and demonstrate accurate detection of sleep timing, achieving mean sleep onset and offset errors of 6.3 and 7.4 minutes, respectively, along with a Total Sleep Time error of approximately 42 minutes. These results indicate that the pipeline can reliably identify primary sleep periods while maintaining practical accuracy for sleep duration estimation.

Calibrated once on a public dataset and applied unchanged across users, the method demonstrates strong generalization without per-user tuning, making it suitable for deployment on emerging wearable platforms. We further validate this through a cross-device study using the WeBe Band in a co-wear setting with a commercial ActiGraph device.

While effective for sleep period detection, the approach remains limited in capturing short wake events, as reflected in higher WASO error. Future work includes improving wake sensitivity, incorporating additional sensing modalities such as EDA and PPG, exploring adaptive calibration strategies, and conducting larger-scale studies to further evaluate robustness in real-world settings.



\bibliographystyle{ACM-Reference-Format}
\bibliography{ref.bib}

@article{atkinson2007relationships,
  title={Relationships between sleep, physical activity and human health},
  author={Atkinson, Greg and Davenne, Damien},
  journal={Physiology \& behavior},
  volume={90},
  number={2-3},
  pages={229--235},
  year={2007},
  publisher={Elsevier}
}

@inproceedings{fang2024validation,
  title={Validation of webe band during physical activities},
  author={Fang, Ruijie and Hang, Sally and Zhang, Ruoyu and Fang, Chongzhou and Rafatirad, Setareh and Hostinar, Camelia and Homayoun, Houman},
  booktitle={2024 IEEE 20th International Conference on Body Sensor Networks (BSN)},
  pages={1--4},
  year={2024},
  organization={IEEE}
}

@inproceedings{zhang2024introducing,
  title={Introducing we-be band: an end-to-end platform for continuous health monitoring},
  author={Zhang, Ruoyu and Fang, Ruijie and Orooji, Mahdi and Homayoun, Houman},
  booktitle={2024 46th Annual International Conference of the IEEE Engineering in Medicine and Biology Society (EMBC)},
  pages={1--5},
  year={2024},
  organization={IEEE}
}

@article{shao2025know,
  title={Know me by my pulse: Toward practical continuous authentication on wearable devices via wrist-worn ppg},
  author={Shao, Wei and Liang, Zequan and Zhang, Ruoyu and Fang, Ruijie and Miao, Ning and Kourkchi, Ehsan and Rafatirad, Setareh and Homayoun, Houman and Fang, Chongzhou},
  journal={arXiv preprint arXiv:2508.13690},
  year={2025}
}

@article{liang2025rapid,
  title={Rapid Adaptation of SpO2 Estimation to Wearable Devices via Transfer Learning on Low-Sampling-Rate PPG},
  author={Liang, Zequan and Zhang, Ruoyu and Shao, Wei and Kourkchi, Ehsan and Rafatirad, Setareh and Homayoun, Houman and others},
  journal={arXiv preprint arXiv:2509.12515},
  year={2025}
}

@inproceedings{liang2025generalizable,
  title={Generalizable Blood Pressure Estimation from Multi-Wavelength PPG Using Curriculum-Adversarial Learning},
  author={Liang, Zequan and Zhang, Ruoyu and Shao, Wei and Nejad, Mahdi Pirayesh Shirazi and Kourkchi, Ehsan and Rafatirad, Setareh and Homayoun, Houman},
  booktitle={2025 IEEE 21st International Conference on Body Sensor Networks (BSN)},
  pages={1--4},
  year={2025},
  organization={IEEE}
}

@inproceedings{shao2025self,
  title={Self-Supervised and Topological Signal-Quality Assessment for Any PPG Device},
  author={Shao, Wei and Zhang, Ruoyu and Liang, Zequan and Kourkchi, Ehsan and Rafatirad, Setareh and Homayoun, Houman},
  booktitle={2025 IEEE 21st International Conference on Body Sensor Networks (BSN)},
  pages={1--4},
  year={2025},
  organization={IEEE}
}

@article{acebo2006actigraphy,
  title={Actigraphy.},
  author={Acebo, Christine and LeBourgeois, Monique K},
  journal={Respiratory care clinics of North America},
  volume={12},
  number={1},
  pages={23--30},
  year={2006}
}

@article{PhysioNet-mmash-1.0.0,
  author = {Rossi, Alessio and {Da Pozzo}, Eleonora and Menicagli, Dario and Tremolanti, Chiara and Priami, Corrado and Sirbu, Alina and Clifton, David and Martini, Claudia and Morelli, David},
  title = {{Multilevel Monitoring of Activity and Sleep in Healthy People}},
  journal = {{PhysioNet}},
  year = {2020},
  month = jun,
  note = {Version 1.0.0},
  doi = {10.13026/cerq-fc86},
  url = {https://doi.org/10.13026/cerq-fc86}
}

@article{cole1992automatic,
  title={Automatic sleep/wake identification from wrist activity},
  author={Cole, Roger J and Kripke, Daniel F and Gruen, William and Mullaney, Daniel J and Gillin, J Christian},
  journal={Sleep},
  volume={15},
  number={5},
  pages={461--469},
  year={1992},
  publisher={Oxford University Press}
}

@article{sadeh1994activity,
  title={Activity-based sleep-wake identification: an empirical test of methodological issues},
  author={Sadeh, Avi and Sharkey, M and Carskadon, Mary A},
  journal={Sleep},
  volume={17},
  number={3},
  pages={201--207},
  year={1994},
  publisher={Oxford University Press}
}

@article{tudor2014fully,
  title={Fully automated waist-worn accelerometer algorithm for detecting children’s sleep-period time separate from 24-h physical activity or sedentary behaviors},
  author={Tudor-Locke, Catrine and Barreira, Tiago V and Schuna Jr, John M and Mire, Emily F and Katzmarzyk, Peter T},
  journal={Applied physiology, nutrition, and metabolism},
  volume={39},
  number={1},
  pages={53--57},
  year={2014},
  publisher={NRC Research Press}
}

@article{marino2013measuring,
  title={Measuring sleep: accuracy, sensitivity, and specificity of wrist actigraphy compared to polysomnography},
  author={Marino, Miguel and Li, Yi and Rueschman, Michael N and Winkelman, John W and Ellenbogen, Jeffrey M and Solet, Jo M and Dulin, Hilary and Berkman, Lisa F and Buxton, Orfeu M},
  journal={Sleep},
  volume={36},
  number={11},
  pages={1747--1755},
  year={2013},
  publisher={Oxford University Press}
}

@article{de2019wearable,
  title={Wearable sleep technology in clinical and research settings},
  author={De Zambotti, Massimiliano and Cellini, Nicola and Goldstone, Aimee and Colrain, Ian M and Baker, Fiona C},
  journal={Medicine and science in sports and exercise},
  volume={51},
  number={7},
  pages={1538},
  year={2019}
}

@article{freedson2005calibration,
  title={Calibration of accelerometer output for children},
  author={Freedson, Patty and Pober, David and Janz, Kathleen F},
  journal={Medicine \& Science in Sports \& Exercise},
  volume={37},
  number={11},
  pages={S523--S530},
  year={2005}
}

@article{matthew2005calibration,
  title={Calibration of accelerometer output for adults},
  author={Matthew, Charles E},
  journal={Medicine \& Science in Sports \& Exercise},
  volume={37},
  number={11},
  pages={S512--S522},
  year={2005}
}

@misc{webe_web,
  title={We-Be Band – Healthetile},
  author={HealtheTile},
  year={2026},
  howpublished = "\url{https://healthetile.io/product/we-be-band/}",
  note = "[Online; accessed May 2026]"
}

@misc{actigraph_web,
  title={ActiGraph LEAP | Ametris Wearable Devices},
  author={Ametris},
  year={2026},
  howpublished = "\url{https://ametris.com/actigraph-leap}",
  note = "[Online; accessed May 2026]"
}

\end{document}